\documentclass[manuscript]{acmart}

\AtBeginDocument{%
  \providecommand\BibTeX{{%
    \normalfont B\kern-0.5em{\scshape i\kern-0.25em b}\kern-0.8em\TeX}}}

\copyrightyear{2023} 
\acmYear{2023} 
\setcopyright{rightsretained} 
\acmConference[TAS '23]{First International Symposium on Trustworthy Autonomous Systems}{July 11--12, 2023}{Edinburgh, United Kingdom}
\acmBooktitle{First International Symposium on Trustworthy Autonomous Systems (TAS '23), July 11--12, 2023, Edinburgh, United Kingdom}\acmDOI{10.1145/3597512.3599715}
\acmISBN{979-8-4007-0734-6/23/07}

\graphicspath{{figures}}
\usepackage{hyperref}
\usepackage{url}
\usepackage{blindtext}
\usepackage{xcolor}

\begin{document}

\title{Perceived Trustworthiness of Natural Language Generators}

\author{Beatriz Cabrero-Daniel}
\email{beatriz.cabrero-daniel@gu.se}
\orcid{0000-0001-5275-8372}
\email{webmaster@marysville-ohio.com}
\affiliation{%
  \institution{University of Gothenburg}
  \streetaddress{Hörselgången 11}
  \city{Göteborg}
  \country{Sweden}
  \postcode{41756}
}
\author{Andrea Sanagustín Cabrero}

\renewcommand{\shortauthors}{Cabrero-Daniel and Sanagustín}

\begin{abstract}
Natural Language Generation tools, such as chatbots that can generate human-like conversational text, are becoming more common both for personal and professional use. However, there are concerns about their trustworthiness and ethical implications.
The paper addresses the problem of understanding how different users (e.g., linguists, engineers) perceive and adopt these tools and their perception of machine-generated text quality. It also discusses the perceived advantages and limitations of Natural Language Generation tools, as well as users' beliefs on governance strategies.
The main findings of this study include the impact of users' field and level of expertise on the perceived trust and adoption of Natural Language Generation tools, the users' assessment of the accuracy, fluency, and potential biases of machine-generated text in comparison to human-written text, and an analysis of the advantages and ethical risks associated with these tools as identified by the participants. Moreover, this paper discusses the potential implications of these findings for enhancing the AI development process.
The paper sheds light on how different user characteristics shape their beliefs on the quality and overall trustworthiness of machine-generated text. Furthermore, it examines the benefits and risks of these tools from the perspectives of different users.
\end{abstract}





\maketitle


\section{Introduction}


Natural Language Generation (NLG) tools, such as chatbots, are becoming more common both for personal and professional use. They have already been widely accepted in various industries, such as customer service or translation, and their popularity is soaring after the public release of OpenAI's ChatGPT\footnote{Visit \url{chat.openai.com/chat}}. However, despite the growing adoption of NLG tools, some users are still doubtful: some do not trust Artificial Intelligence (AI) and some even fear that it will replace humans at the workplace.

As the adoption rates continue to grow, it is important to investigate how different groups of users perceive and use these tools. Although their developers may have confidence in the robustness of NLG tools, it is unclear how the general population perceives them. In order to address the concerns of users and build trust for NLG applications, it is crucial to ensure that these AI tools are not just robust, but also lawful and ethical~\cite{AIact}. Without trust, it is unlikely that NLG tools will be adopted in fields that could benefit the most from their applications, such as healthcare or education~\cite{Bates19,Tan2023}.

It is therefore important to understand and address any concerns users may have regarding the accuracy, fairness, and ethical implications of NLG tools~\cite{Ogunniye2021}. Moreover, researchers need to develop effective strategies to communicate the benefits of NLG tools while also addressing any perceived limitations and ethical concerns that may arise. Ultimately, these efforts aim to guide development and deployment of AI systems that benefit of users, while also taking into account the concerns and needs of all other stakeholders. Only then we can fully achieve the potential of NLG tools to transform industries and improve our daily lives.

The present paper reports a quantitative and qualitative study on the effect of different user characteristics and beliefs on perceived trustworthiness of Natural Language Generation tools. Specifically, the research questions that this paper aims to answer are:

\begin{itemize}
    \item \textbf{RQ1.} How do user characteristics affect perception of trustworthiness and adoption of NLG tools?
    \item \textbf{RQ2.} How is the quality of machine-generated texts perceived, including the ability to pass as human authored?
    \item \textbf{RQ3.} What are the advantages and limitations of NLG tools according to different sets of users? What governance strategies do they recommend?
\end{itemize}

Sections~\ref{sec:rq1} and~\ref{sec:rq2} discuss the responses of a survey on trustworthiness and quality of machine-generated text. The survey responses on advantages and limitations of NLG, and their beliefs about its ethical and social implications, are summarised in Section~\ref{sec:rq3}.

\section{Background}

\textbf{Human-computer interaction} (HCI) focuses on the role of human users and their contexts in the design of interactive systems~\cite{dix2003human}, including Autonomous Systems (AS), and on the promotion of principles such as usability, accessibility, consistency, or generalisability. As such, HCI is a complex and rapidly changing field: for example, only recently we have been able to engage in ubiquitous and seamless interactions with smart speakers or chatbots.
Research in HCI studies how to identify potential issues and ensure that the system meets the needs and expectations of end-users~\cite{dix2003human,Ogunniye2021}. For instance, previous research on natural language interaction focuses on usability and user judgement~\cite{bleu2002,Bentivogli2016, Salminen2022}. Nevertheless, with the rise of AI NLG tools such as chatbots, further research is needed to address the new interaction paradigms, especially focusing on trustworthiness.

\textbf{Natural Language Processing} (NLP) is a field of Machine Learning that deals with linguistics using probabilistic and statistical techniques to determine the likelihood of word sequences occurring in a sentence~\cite{Singh21}. These models are useful for both Natural Language Understanding, which aims to translate user input into formal representations, and for Natural Language Generation (NLG), used to generate dialect utterances for NLG tasks like automatic text translation or automatic insight reporting of large datasets~\cite{Singh21}. For example, NLG has been since long used in bilingual weather report generation systems~\cite{bateman1997}, or in the production of personalised healthcare materials, addressing the professionals' limited time, and resulting in more specific and effective communication of patient information~\cite{Cawsey1997}. 

\textbf{Language models} typically recreate the sequential nature of language using Recurrent Neural Networks~\cite{Singh21}. Lately, foundation models, based on large scale deep neural networks trained on extensive self-supervised data~\cite{gpt3,Foundation2022}, are surpassing previous attempts due to their general intelligence. These complex methods also enable adaptive terminology and style to meet specific readership needs~\cite{Cawsey1997}. Nevertheless, text quality requires balancing fluency, coherence, and flexibility which is challenging and often requires human oversight and post-editing~\cite{Cawsey1997}. Moreover, these colossal NLG tools, such as GPT-3, require being trained on GPU clusters for days or weeks, which makes them costly resource-wise~\cite{Singh21}. To meet the growing demand for NLP and NLG on edge devices, compressed models such as ChatGPT\footnote{Visit \url{chat.openai.com}} have been introduced. As discussed in this paper, these NLG tools can't be trusted out-of-the-box.


\textbf{Chatbots}, the focus of this work, are language models embedded in everyday devices, such as smartphones or smart speakers, to simulate human conversation through text or speech~\cite{Chaix19}. Chatbots can improve the access to information in many fields, e.g., healthcare~\cite{Bates19,Chaix19,Bharti20}, legal~\cite{Queudot20}, customer service~\cite{Queudot20,Bates19,Salminen2022}, or even e-recruitment~\cite{Ogunniye2021}. The rise in popularity of chatbots such as ChatGPT, which reached more than 1 million users in the first 5 days after its release, pose questions related to HCI issues~\cite{chatgpt}.

\textbf{Machine Translation} (MT) is an important application of NLP in an increasingly globalised world~\cite{Schiebinger2013,Singh21}. Even though the error rates of publicly available MT systems are still high, recent improvements based on Deep Learning are leading to higher quality results in terms of accuracy, fluency, etc. To train MT systems, text in one language alongside its translation in another language is needed, together with a corpus grammar of the language being translated into~\cite{Schiebinger2013}. Contrary to publicly available MT tools, Computer Assisted Translation (CAT) tools are designed to support and speed up the work of professional translators that rely on a translation database used to retrieve suggestions from previously translated documents~\cite{Bentivogli2016}.  

\textbf{Ensuring AI trustworthiness}, including chatbots, is the goal of a number of legal frameworks.
According to the EU AI Act, a proposed European Union law, all AI systems that pose a significant risk to the health and safety or fundamental rights of persons should be trustworthy, i.e., robust, lawful, and ethical~\cite{Bentivogli2016,AIact}. There are many ways in which NLG tools can become high-risk AI, e.g., producing misleading or inaccurate information, biasing the user against a particular social group, or sharing private information about users. As high-risk AI systems, NLG tools should aim to be trustworthy, but embedding ethical values in NLG is challenging~\cite{Ozkaya2019,Foundation2022,Tan2023}. Classical AI verification and validation mechanisms may not capture ethical values~\cite{Berry2022,Halme2022,Aydemir2018} so the EU AI Act requires human oversight mechanisms~\cite{TAIEU}. Chatbots must therefore clearly inform their diverse sets of users about their characteristics, capabilities and limitations of performance~\cite{Dao2009,Gold2020}. 
While all the necessary control mechanisms and communication channels might be in place, users might have preconceived ideas about AS that can influence their perception of AI trustworthiness. For instance, in some cases humans tend to mistrust AS, especially in safety-critical situations. In other cases, humans tend to show automation bias, e.g., automatically relying or over-relying on the output produced by a chatbot~\cite{AIact,Dao2009}. Moreover, different sets of users might show different levels of trust and interact with AS in very different ways as a result~\cite{dix2003human}.

\textbf{Unfair bias.}
While human chatbot users might be biased against AI due to previous life experiences, chatbots might be biased against some population groups due model flaws, insufficient training, or cyber-attacks~\cite{AIact}. To be ethical, NLG needs to avoid unfair bias, especially in critical areas such as education, employment, healthcare, or law enforcement~\cite{doshi2017accountability,Bates19,AIact}. Some of the techniques used to avoid bias in NLG, for instance regarding gender, consists on using rich contextual information to improve the readability and quality of machine translations~\cite{Schiebinger2013}. NLG systems can also exploit linguistic theories about where pronouns can be used and what they can be used to refer to, to automatically choose between using a pronoun or a full noun phrase at a particular point in a text~\cite{Cawsey1997}.

\section{Methodology} \label{sec:method}


Given our aim to assess the trust on natural language generators by different sets of users and to establish causal relationships between trust-related variables (e.g., expertise, adoption, explainability, etc.), we designed an online survey\footnote{Take the survey: \url{https://forms.office.com/e/Z4pGzwqmwX}}created using \textit{Microsoft Forms}, in which respondents answered a series of mostly structured questions.
The target population for the survey were professionals related to either modern languages and linguistics (Group 1) or software design and development (Group 2), including AI and NLP tools. 
Participants were recruited by direct contact and by broadcasting the survey among \textit{LinkedIn} targeted groups. Incentives were not offered but a 1-minute-long video\footnote{Watch the video: \url{https://play.gu.se/media/0_wuxsjuwh}} discussing trustworthy AI was used as a ``call to action'' to improve voluntariness. 
This resulted in a sample of $N=77$ respondents. The number of users of AI tools for NLG is rapidly growing; nevertheless we believe this sample is sufficient to get representative answers that correlate with general trends.

Restricted experts refer to participants that answered 1, 2 or 3 to the question ``From 1 (still studying/recently graduated) to 5 (expert in the field), what is your level of expertise in your field (according to your own perception)?'' We expect these participants to hold a narrow and intuitively-based interpretation of their field. On the other hand, we expect ``extended'' experts to hold a much broader vision and enhanced knowledge. 
Among the participants in the survey, 54\% are non-experts non-users, 27\% are non-experts adopters, 12\% are experts non-users, and 5\% are experts adopters. The participants demographics are detailed in Table~\ref{tab:demographics} where: 
\begin{itemize}
    \item \textbf{Group 1}: linguists, translators, interpreters, or related.
    \item \textbf{Group 2}: participants whose field of expertise is Artificial Intelligence, Natural Language Processing, Computer Science, Software Engineering, or similar fields.
    \item \textbf{Group 3}: participants in any other field (not reported).
\end{itemize}

We piloted the online survey with participants from our contact set. The pilots allowed us to estimate and improve response rates by refining the questions and the ``call to action''. No data from the pilots were retained in the final data set and pilot participants did not participate in the final round. 
The response rate for the survey was approximately 50\% for the directly contacted participants (approximately 90\% of the total respondents). The percentage was significantly lower, closer to 30\%, for respondents in Group 2. 

\begin{table}[h]
\centering
\begin{tabular}{@{}lccc@{}}
\hline
\textbf{Group by} & \textbf{Restricted} & \textbf{Extended} & \textbf{Experiment} \\
\textbf{field} & \textbf{experience} & \textbf{experience} & \textbf{participation} \\ \hline
Group 1 & 31 & 3 & 91\% \\
Group 2 & 17 & 7 & 87\% \\
Group 3 & 15 & 4 & 68\% \\ \hline
\end{tabular}
\caption{Expertise of survey participants in their fields.}
\label{tab:demographics}
\end{table}


The online survey consisted of (i) questions regarding demographics, (ii) Likert scale questions to assess respondents' perception of trust-related concepts, and (iii) open-ended questions asking for experienced causes and consequences of positive and negative affect for said tools. The questions are mapped to the research questions but appropriate wording and descriptions, as learnt during the pilot, were needed to adapt to the non-expert audience. 

Additionally, the survey had an optional section aimed at evaluate the ability to detect machine-generated text and translations. It was restricted to voluntary participants with a C2 level, or native, in the English, French, German, Spanish, or Swedish languages. The participants were asked to rate a number of sentences on a Likert scale of 1 to 5 from low to high likelihood of human authorship. The experiment also contained a question where participants were asked to order four sentences depending on their likelihood to have been written by humans. The hypothesis tested in this experiment is whether there is a significant difference between the likelihood of participants identifying sentences written by humans compared to those machine-generated or machine-translated. Additionally, the experiment explored whether certain factors such as the language used or the field of expertise of participants affect the likelihood of identifying the origin of the sentence.


On the other hand, data analysis was carried out to identify emerging concepts from the replies to the open-ended questions. Before analysis, we cleaned the data set by removing empty replies and replies not relevant to the research questions. Then, a thematic analysis was conducted out by the second author to code the responses in light of the research questions and then revised by the first author in a collaborative session. An \textit{a priori} coding scheme was created by identifying  emergent themes. Parent themes are:
\begin{itemize}
    \item General beliefs about AI-generated text
    \item Advantages of natural language generators
    \item Future functionalities with positive impacts
    \item Future negative impacts and ethical risks
    \item Regulation and mitigation strategies
\end{itemize}

The statistical analysis of the replies to the close-ended questions, for all the sections of the survey, is provided in Sections~\ref{sec:rq1} and~\ref{sec:rq2}, the responses of the open-ended questions are analysed in~\ref{sec:rq3}, whilst the complete set of questions and non-confidential aggregated survey responses are provided as \textit{Supplementary Materials}.

\subsection{Threats to validity}


Several potential threats to validity concern the study sample:

On the one hand, although we applied purposeful sampling to send the survey, and direct contacts represent the majority of the respondents, the survey was available online and anonymous, making it impossible to verify the sources. Moreover, 42\% of respondents reported being students or having recently graduated. In addition, our sample is not diverse in terms of gender in Group 2, a known bias within the Software Engineering community. 

Another limitation of this study could be the use of terminology from the AI Act, which may not have been suitable for linguists who do not have expertise in software engineering and business management. To address this issue, the survey's pilot phase was instrumental in overcoming these difficulties by providing appropriate definitions for the terms used in the questions. Despite these efforts, some of the survey questions may still have been challenging for linguists to comprehend due to unfamiliar or new concepts, e.g., explainability, auditability, etc.

It is possible that participants in this study were biased against AI tools for NLG due to the confusion spread by the media at the time of the survey, which was conducted only two months after the release of ChatGPT for the general public.

Whether any remaining bias in the sample has an impact on the perceived trustworthiness of NLG tools is an open question that requires replication with other samples to be addressed. 

\section{Perception of trustworthiness and adoption of NLG tools} \label{sec:rq1}

One key factor that may impact users' perception of NLG tools is their field of work and level of expertise. Research has shown that individuals with greater expertise may have lower expectations for the performance of these tools, and may be more critical of their limitations. Therefore, our first research question seeks to explore how these user characteristics affect the perception of trustworthiness and adoption of NLG tools.

\subsection{Adoption and non-usage of NLG tools}

In the survey, participants participants from all fields reported whether they were aware of the existence of or had even used some AI tools. The most commonly known and used tool was machine translators, with 97\% of participants knowing and using them. Chatbots, such as ChatGPT, were also relatively well-known and used, with 77\% of participants indicating they had experience with this type of tool. AI image creators and CAT (computer-assisted translation) software were less commonly known and used, with 45\% and 35\% of participants reporting familiarity, respectively.

Adopters are respondents that know at least two machine translators or text generation tools and use at least one in their work. As can be seen in Table~\ref{tab:adoptiongroups}, the adoption of such tools varies significantly depending on the field and expertise level of the participants. Among non-experts in Group 1, 32\% are adopters, while in Group 2, 59\% are adopters. Interestingly, none of the experts in Group 1 are adopters, indicating that they may rely less on these tools due to their expertise in the language. In contrast, 43\% of experts in Group 2 are adopters. Moreover, there are differences in the willingness to use NLG tools in work depending on the participants' field of work and expertise level (see Table~\ref{tab:knowanduse}). These differences might be explained by the different beliefs of these groups, outlined in Section~\ref{sec:rq2} and~\ref{sec:rq3}. 

\begin{table}[h]
\centering
\begin{tabular}{lcc}
\hline
\textbf{Group by} & \textbf{Restricted} & \textbf{Extended} \\
\textbf{field} & \textbf{experience} & \textbf{experience} \\
\hline
Group 1 & 32\% & 0\% \\
Group 2 & 59\% & 43\% \\
Group 3 & 7\% & 25\% \\
\hline
\end{tabular}
\caption{Adoption of NLG tools by field and expertise level.}
\label{tab:adoptiongroups}
\end{table} 

\begin{table}[h]
    \centering
    \begin{tabular}{lcc}
    \hline
         \textbf{Group by} & \textbf{Machine} & \textbf{Chatbots, language}  \\
        \textbf{field} & \textbf{translators} & \textbf{generation tools}  \\
    \hline
    Group 1 & 100\%, 79\% & 76\%, 35\% \\
    Group 2 & 100\%, 26\% & 100\%, 73\% \\
    Group 3 & 89\%, 57\% & 63\%, 15\% \\
    \hline
    \end{tabular}
    \caption{Percentage of awareness of existence (first percentage) and usage of NLG tools in work (second percentage).}
    \label{tab:knowanduse}
\end{table}

A surprising result, reported in Table~\ref{tab:knowanduse}, is that more than a third of respondents in Group 1 (35\%) reported using chatbots and other NLG tools in their work, while almost three quarters of respondents in Group 2 (73\%) said the same. This result might suggest that AI and NLG tools are rapidly becoming acceptable to be used in work, even among language-related professionals. It would be interesting to investigate further which workflows are more likely to be adapted using NLG and what oversight mechanisms can be used.

\subsection{Effect of expertise on trust in NLG tools} 

\begin{figure*}[h]
    \centering
    \includegraphics[width=\linewidth]{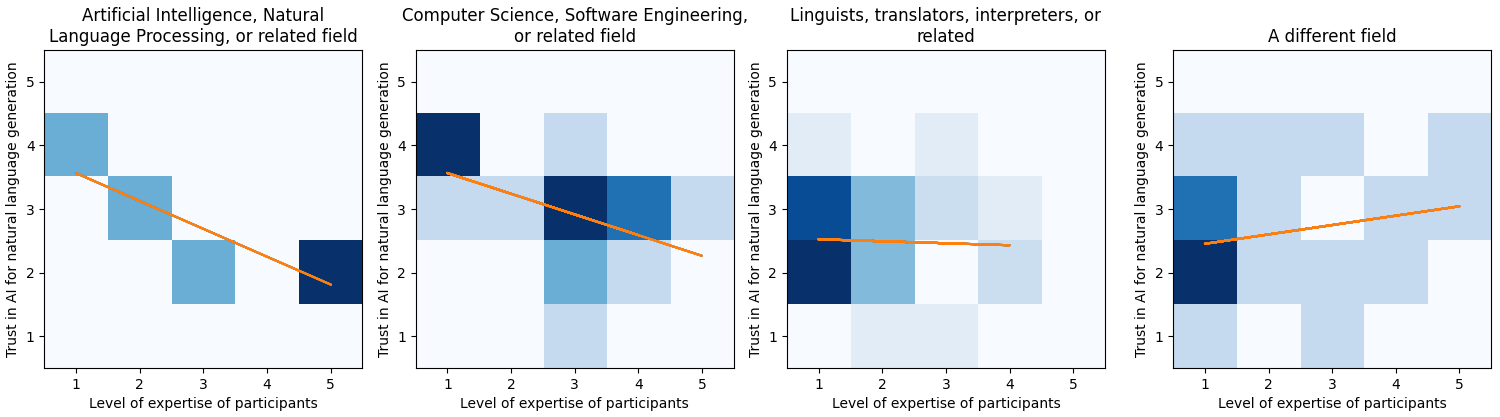}
    \caption{Expertise effect on trust in Natural Language Generation tools per field (blue) and linear regression analysis (orange).}
    \label{fig:correlationexpertiseinAItrust}
\end{figure*}

As seen in Table~\ref{tab:adoptiongroups}, the field and expertise level are related to different beliefs about NLG tools. This result is complemented by Figure~\ref{fig:correlationexpertiseinAItrust}, that aggregates the responses to the question ``From 1 to 5, how much would you say you trust Artificial Intelligence (AI) tools for Natural Language Generation?''.

This analysis suggests that there is a difference in trust in NLG tools based on the level of experience with ICT. Participants with more experience in ICT reported lower levels of trust in NLG tools compared to those with less experience: the correlation between level of expertise and trust in AI tools for NLG for participants working on Group 2 is -0.87. Figure~\ref{fig:correlationexpertiseinAItrust} represents the aforementioned negative correlation between the level of expertise and the trust in AI tools for NLG. This difference in trust levels was not observed in Group 1 nor Group 3 (e.g., the correlation is -0.03 for Group 1). 

Deeper knowledge in software systems may influence participants' beliefs about the trustworthiness of AI tools for NLP. Nevertheless, when answering the question ``From 1 to 5, how much would you say you trust Artificial Intelligence tools for Natural Language Generation?'' the average responses of the groups ``Artificial Intelligence, Natural Language Processing, or related field'' (2.6 $\pm$  0.8), ``Computer Science, Software Engineering, or related field'' (2.8 $\pm$ 0.81), ``A different field'' (2.55 $\pm$ 0.89), ``Linguists, translators, interpreters, or related'' (2.48 $\pm$ 0.7), are not statistically different.

\subsection{Perception of NLG trustworthiness and adoption rates per field of work}

\begin{figure}[h]
    \centering
    \includegraphics[width=.6\linewidth]{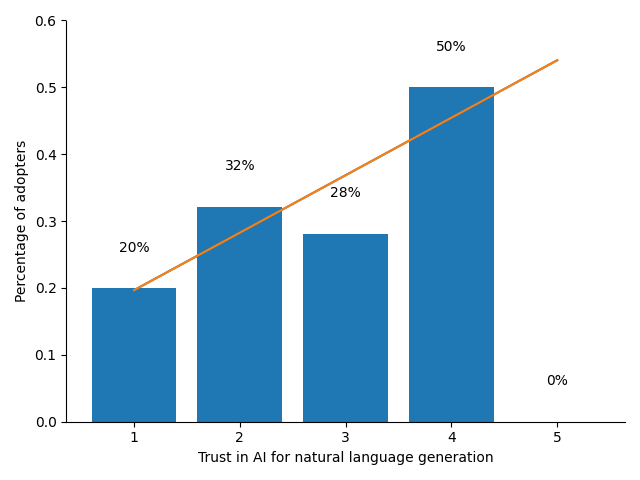}
    \caption{Percentage of adoption of machine translators and NLG tools by participants' trust in AI for NLG.}
    \label{fig:trustonusage}
\end{figure}

It is unclear from the survey responses whether users adopt NLG tools because they trust them or if they trust NLG tools because they have adopted and experienced them. Regardless of the direction of the causative relationship, the correlation between the trust in and adoption of NLG tools is significant, as can be seen in Figure~\ref{fig:trustonusage}.

The analysis of the survey responses shows that participants who selected higher trust scores for NLG tools, on a scale of 1 to 5, were more likely to adopt them. Interestingly, no participant selected the highest rating of 5, indicating that there is still room for improvement in the perceived trustworthiness of NLG tools. This highlights the importance of building trust in AI and NLG tools to increase their adoption for tasks such real-time transcriptions.
These findings relate trust in NLG tools, informed or not, to an increased adoption, and potential misuse, in various industries.


\section{Perception of quality of machine-generated texts} \label{sec:rq2}

An important aspect of NLG tools is the quality of the generated text. Despite advances in recent years, machine-generated text is still prone to contain grammatical errors, unfair biases, or inappropriate comments. Our second research question aims to investigate how different groups of users perceive the quality of machine-generated text, including its ability to pass as human-authored.

\subsection{Compared accuracy, fluency, cultural appropriateness, and overall text quality}

When asked about their perception of machine-written text characteristics such as fluency, most participants agree that humans perform better than text generators. The analysis of their responses shows that:

\begin{itemize}
	\item 59\% of participants think that fluency in AI-generated text is as good, or even better, as in human produced text. 
	\item 33\% of participants think that accuracy in AI-generated text is as good, or even better, as in human produced text.
	\item 31\% of participants think that overall quality in AI-generated text is as good, or even better, as in human produced text.
	\item 20\% of participants think that cultural appropriateness in AI-generated text is as good, or even better, as in human produced text. 
\end{itemize}

\begin{figure*}[h]
    \centering
    \includegraphics[width=\linewidth]{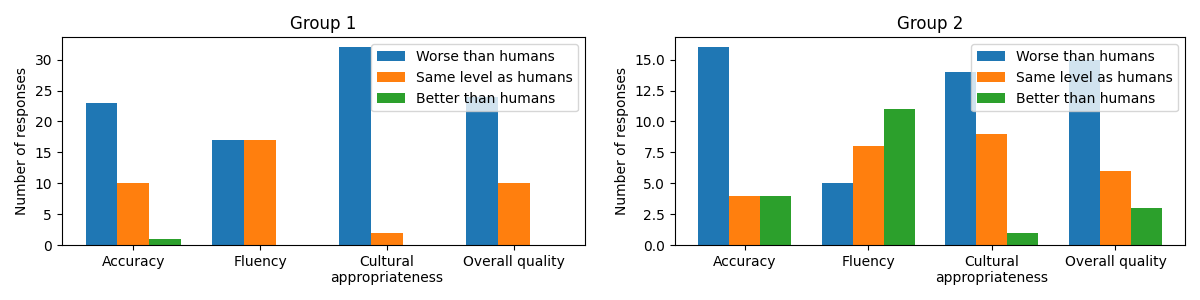}
    \caption{Groups 1 and 2's perception of how machine generated texts and translations compare to those produced by human beings.}
    \label{fig:accuracyfluency}
\end{figure*}

The perception of machine-written text characteristics, though, greatly depends on some user characteristics and usage of these tools. Figure~\ref{fig:accuracyfluency} compares Group 1 and Group 2's perception of a number of text properties, such as accuracy or fluency, of machine generated texts compared to texts produced by human beings. 
The gap between participants in different fields, especially for cultural appropriateness (see Table~\ref{tab:beliefs}), could be explained by participants in Group 2 being overconfident in their own field of work while Group 1 might be wary of the technology or simply better at recognising AI nuances. Unfortunately, the analysis of the experiment responses, in Section~\ref{sec:experiment}, can't support or reject this claim because contextual information was not provided that would allow participants to evaluate these properties. Further research relating participants' language skills and level of contextual information provided in the analysed texts would be required to support this claim.

\begin{table}[h]
\centering
\begin{tabular}{lcccc}
\hline
\textbf{\begin{tabular}[c]{@{}l@{}}Group by\\ field\end{tabular}} & \textbf{Accuracy} & \textbf{Fluency} & \textbf{\begin{tabular}[c]{@{}c@{}}Cultural\\ appropriateness\end{tabular}} & \textbf{\begin{tabular}[c]{@{}c@{}}Overall\\ quality\end{tabular}} \\
\hline
Group 1 & 32\% & 50\% & 6\% & 29\% \\
Group 2 & 33\% & 76\% & 43\% & 38\% \\
Group 3 & 35\% & 47\% & 18\% & 24\% \\
\hline
\end{tabular}
\caption{Percentage of participants per field that think AI-generated text is as good, or better, than human produced text for a number of properties.}
\label{tab:beliefs}
\end{table}

Regardless of the cause of this gap, these results highlight the importance of involving individuals with different backgrounds, perspectives, and experiences in the AI development process to design balanced goals for NLG tools, in terms of text characteristics, to meet the needs of all users.

\subsection{Unfair biases and compliance with privacy and data protection laws} \label{sec:beliefs}

The survey found that 19\% of participants were unsure if there are differences in machine translation performance for certain languages, language pairs, or specific domains. Among the rest of the participants, 90\% said that there are differences in performance for certain pairs or domains. Only a small number of adopters indicated that they did not believe there were differences.
The belief that there are certain languages, language pairs, or specific domains for which machine translation tools perform better or worse is shared by 91\% of adopters and 88\% of non-users. Participant P15 expressed the concern that excessive reliance on machine translators and NLG tools could lead to a decrease in efforts to study different languages from around the world. 

Whilst a minority of participants mentioned in the open-ended questions that the inherent biases present in our world can influence the AI models, only 6\% of participants think that there is no bias in NLG and that they are as good as humans, or better, in avoiding gender bias. On the other hand, 84\% of participants agreed that machine-generated texts are gender biased. Out of them, 43\% think that even though NLG is inherently gender biased, it is as good as humans, or better, in avoiding gender bias. 

On biased machine-generated content, P34 highlights the importance of gender accuracy in machine-generated texts to make machine-generated texts closer to human-written ones. P48 points out the risk of a positive feedback loop, where ``the bias will amplify'' due to inadvertently training AI systems using inherently biased NLG content. This can happen for any form of bias in AI systems, including ``racial and gender bias'' (P48). In relation to biases, P49 mentioned the need to ``understand the emotions of the person the AI is talking to to craft intentional, coherent, and polite answers''. 

\begin{figure}[h]
    \centering
    \includegraphics[width=.6\linewidth]{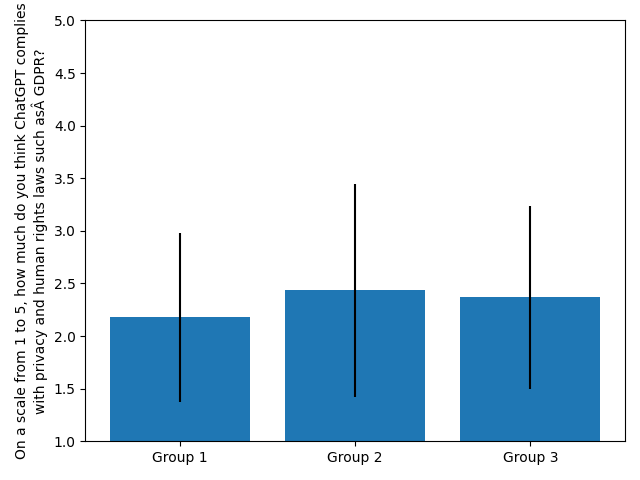}
    \caption{Perception of ChatGPT's compliance with privacy and human rights laws.}
    \label{fig:gdpr}
\end{figure}

When asked whether ChatGPT complies with privacy and human rights laws such as GDPR, several participants expressed concerns about the use of personal information and data protection laws. P59 said that they are still worried about the use of their personal information. P33 shared the same concern and emphasised the importance of ``having a clause in Terms and Services that ensures data won't be sold or used for purposes outside of the task at hand.'' 
As can be seen in Figure~\ref{fig:gdpr}, the means of the three groups are not statistically different when answering the question ``On a scale from 1 to 5, how much do you think ChatGPT complies with privacy and human rights laws such as GDPR?'' Similarly, the perception of compliance with such laws is also not statistically different between adopters and non-users of NLG tools.

\subsection{Perception of likelihood of human authorship: experiment results} \label{sec:experiment}


In the experiment, described in Section~\ref{sec:method}, participants were asked to rate sentences on a Likert scale of 1 to 5 from low to high likelihood of human authorship. Some of the sentences were machine-generated or machine-translated. The experiment also contained a question where participants were asked to four sentences depending on the perceived likelihood of being written by humans. 

The data collected from these ratings allows for comparisons between the different types of sentences presented (human-written, ChatGPT-written, machine-translated). Additionally, the experiment explored whether certain factors such as the language used (see Table~\ref{tab:scoresperlang}) or the field of expertise (see Table~\ref{tab:scorespergroup}) of participants influence the likelihood of identifying the origin of the sentence.

\begin{table}[h]
\begin{tabular}{lcc}
\hline
\textbf{Language} & \textbf{Human-written} & \textbf{ChatGPT-written} \\ \hline
English & 2.87 $\pm$ 1.26 & 3.36 $\pm$ 1.01 \\
Swedish & 3.50 $\pm$ 1.50 & 3.67 $\pm$ 1.25 \\
Spanish & 2.71 $\pm$ 1.20 & 3.43 $\pm$ 1.26 \\
German & 3.33 $\pm$ 1.25 & 3.50 $\pm$ 1.26 \\
French & 5.00 $\pm$ 0.00 & 4.00 $\pm$ 0.82 \\ \hline
\end{tabular}
\caption{Average scores for human-written and ChatGPT-written text per language}
\label{tab:scoresperlang}
\end{table}

Interestingly, as reported in Table~\ref{tab:scorespergroup}, the average score (on a scale from 1 to 5) for human-written text according to participants in Group 1 is lower than the score given by participants in Group 2 and Group 3. Meanwhile, the average score for ChatGPT-written text according to participants in Group 1 is slightly higher than the score given by Group 2 and Group 3 (3.30 $\pm$ 0.97).

The average scores for ChatGPT-written text per group and per language were also assessed, as reported in Table~\ref{tab:scoresperlang} and in the \textit{Supplementary Materials}. The analysis shows that Group 2 and Group 3 had the highest average score for German (3.83 $\pm$ 1.34). Overall, though, the average score difference across languages is not statistically significant, neither for human-written text nor for ChatGPT-written text

\begin{table}[h]
\begin{tabular}{lcc}
\hline
\textbf{Field} & \textbf{Human-written} & \textbf{ChatGPT-written} \\ 
 \hline
Group 1 & 2.20 $\pm$ 0.40 & 3.47 $\pm$ 1.09 \\
Group 2 & 3.00 $\pm$ 1.15 & 3.33 $\pm$ 0.88 \\
Group 3 & 3.00 $\pm$ 2.00 & 3.00 $\pm$ 1.15 \\ 
\hline
\end{tabular}%
\caption{Average scores (on a scale from 1 to 5) for human-written and ChatGPT-written text according to participants in different groups.}
\label{tab:scorespergroup}
\end{table}

The use of different spellings within German-speaking participants may have influenced their choices when deciding which sentences appeared to be written by humans. This might have biased the participants' responses, as pointed out by the participants in the pilots. The authors, though, decided not to modify the produced texts before participant assessment in the experiment. Future research, in close collaboration with expert linguists, could be conducted how different spellings and NLG styles might affect the participants perception of the generated text across languages.

\section{NLG tools: Limitations, advantages, ethical risks, and governance} \label{sec:rq3}

The survey findings indicate that participants' attitude towards NLG tools is influenced by their background, such as their field of expertise. For example, participants from Group 2 were generally more positive towards the technology and its potential applications. This trend was evidenced in both the structured and open-ended questions, as reported in Sections~\ref{sec:rq1} and~\ref{sec:rq2}. 

It is important to understand and clearly explain the differences in perception of the advantages and limitations of NLG tools, as well as the identified risks in order to develop appropriate regulations and guidelines. Our third research question therefore aims to explore the ethical and governance concerns according to different sets of users, and their thoughts on regulation strategies.

Section~\ref{sec:explain} moves on to cover the responses to the open-ended questions regarding the explainability of NLG tools, while Section~\ref{sec:advantages} and Section~\ref{sec:risks} respectively report the advantages and risks of NLG tools according to the survey respondents.

\subsection{Explainability of NLG tools} \label{sec:explain}

An explainable AI system provides clear and understandable explanations to their users of its decision-making process and outputs. Transparency and explainability are necessary to allow humans to understand and trust the system's decisions.

Table~\ref{tab:fieldexplain} displays the results of a survey investigating the effect of field on perception of ChatGPT's explainability. The results show that approximately a third of survey participants reported believing that ChatGPT clearly explains its decision-making process to its intended users, i.e., its users. Group 1 had the lowest percentage (26\%) of respondents who believed that ChatGPT was explainable to its intended audience. On the other hand, Group 2 had the highest percentage (33\%) of respondents who believed that ChatGPT was explainable to its developers. This suggests that Group 1 might find it difficult to understand how the system works, while Group 2 might have a better understanding of the needs of AI creators, engineers, and developers who were building ChatGPT.

\begin{table}[h]
    \centering
    \begin{tabular}{lccc}
    \hline
    \textbf{Group by} & \textbf{Users (intended} & \textbf{Creators, engineers,} & \textbf{3rd party} \\
    \textbf{field} & \textbf{audience)} & \textbf{and developers} & \textbf{auditors} \\
    \hline
    Group 1 & 26\% & 14\% & 5\% \\
    Group 2 & 33\% & 33\% & 14\% \\
    Group 3 & 26\% & 15\% & 5\% \\
    \hline
    \end{tabular}
    
    \caption{Effect of field on perception of ChatGPT's explainability}
    \label{tab:fieldexplain}
\end{table}

Overall, the results of the survey show that participants' expectations for explainability in chatbots are low, despite the importance of this requirement for trustworthy AI~\cite{AIact}. This finding suggests that measures should be put in place to convey the ability of AI to express its limitations to its users.

\subsection{Advantages and desired functionalities} \label{sec:advantages}

This section reports the advantages of NLG tools that participants could identify or foresee, and their importance in a globalised world. For instance, some of the participants highlighted facilitating communication between people who speak different languages, especially in business, and saving time and resources while adapting to diverse user needs by automatically generating content. 

When asked about the advantages of these tools that outweigh the potential negative effects as of now, participants reported their ability to perform tedious tasks quickly and efficiently (P59), speed up work processes (P53), and increase accessibility by providing fast and cheap language services (P64). Other respondents highlighted the tools' potential for improving communication between different cultures through faster and more efficient translations (P48).

The survey participants also shared their thoughts on the features or capabilities that they would like to see in future NLG tools. Responses from Group 1 regarding future functionalities include:
\begin{itemize}
    \item Human oversight for useful yet time-saving results. 
    \item Coherent and polite answers for user emotional state. 
    \item More advanced machine translation that can handle creative texts and context. 
    \item Improved performance and reliability leading to increased reliance on AI for advice in professional settings. 
\end{itemize}

The automation of repetitive or tedious tasks in various fields and types of work was also mentioned by of participants in Group 2. Other responses from Group 2 regarding features or capabilities that they would like to see in future are:
\begin{itemize}
    \item Faster writing, automatic reports, fast explanations, etc. 
    \item Variations for sentences and online sentence tuning. 
    \item The ability to discuss with AI as if with a human, receiving objective and correct answers. 
    \item Text generation and error fixing. 
    \item Broad and varied vocabulary to produce documentation. 
\end{itemize}

These recommendations could further improve NLG tools and their trustworthiness, make them more efficient, and help users in various industries to generate more accurate, relevant and meaningful insights from the data.

\subsection{Ethical risks and mitigation regulations} \label{sec:risks}

While respondents agreed that NLG tools are a revolution in the way we communicate and work, they also highlight the need to address the risks associated with their use through appropriate regulations. This section discusses the ethical issues and potential negative consequences derived from NLG use that participants reported in the open-ended questions. Three of the most commonly mentioned issues are (i) overlooking privacy and intellectual property rights, (ii) automation bias, and (iii) loss of jobs. 

\textbf{Misinformation and plagiarism.}
Several participants expressed concerns about the potential spread of misinformation and unintentional plagiarism due to the usage of NLG tools. Participants highlighted the risks of overlooking intellectual property risks (P14, P36, P53), e.g., ``wrongly attributing quotes to someone'' (P14). P77 also mentions the risk of confirmation bias. Furthermore, there were concerns about accountability and transparency with regards to the generated artefacts, posing a serious threat to privacy laws (P65). Some also noted the possibility of a ``decrease in creativity and an overall trend towards simplistic, repetitive content'' (P14). Despite attempts to mitigate these issues through regulations, participants felt that more needs to be done to address the potential risks associated with the use of NLG tools.

\textbf{Automation bias.}
Participants also mentioned automation bias (P65, P74, P76), i.e., an over-reliance on NLG content. Some respondents hypothesise that users might ``assume ChatGPT is an objective entity'' (P76) and that ``some people might have a bigger trust in AI because they think they are more impartial, but that's not necessarily the case'' (P77). Participants also provide examples of risks related to automation bias such as ``unfiltered information being taken as truthful'' (P53) and ``forgetting how to gather information or recognise misinformation due to frequent use of ChatGPT'' (P64). Complimentarily, P15 mentioned that over-reliance might result in ``less critical thinking, memory development, knowledge and deductive capability, and correlation thinking'', while P59 mentioned the risk of ``humans losing their writing and communication skills''.

\textbf{Loss of jobs.}
On the one hand, participants shared concerns about the impact of NLG tools on human translators' jobs. They worry that ``the use of natural language synthesis might devalue the position of experts'' (P16) and decrease the quality of translations (P21, P33), leading to a lack of job opportunities for qualified human translators (P16, P17). Some express concern that NLG users ``would be sacrificing quality translations for immediate results, making translators lose jobs while having subpar machine translations'' (P33).
On the other hand, participants expressed scepticism about these tools replacing human translators (P16, P34, P38, P51, P74) and suggested conducting more detailed analyses of the ``impact of complex, high-level translation software on society'' (P15). Nevertheless, the majority of those who responded believe that these tools will keep taking up more space within the translation business, leading to ``a tendency towards full automation'' (P36).

\textbf{Mitigation strategies.}
Participants had varying opinions on future regulation and mitigation strategies for NLG and machine translation tools. Some believed that proper control and regulations were necessary for the development and usage of these tools (P36, P52, P34), including the potential misuse of these tools (P14). Several participants suggested only allowing the ``use of these tools in a controlled environment'' to ``generate knowledge and not to the detriment of it'' (P57). Participants agreed on the need for transparency and data protection for NLG users, and the need of using professional advice to limit applications and devices. P49 also suggested implementing warnings or advisories for users ``such as the one in Netflix when it tells you that you have been watching shows for too long''. P15 highlighted the need to conduct more detailed analyses of the impact of complex, high-level translation software on society and P65 went as far as to recommend ``to stop the public release of such tools until such ethical issues are mitigated, since they pose a risk to humanity and [...] human wellbeing.''

\section{Conclusion}

Despite its novelty, ChatGPT has already become a common tool among many individuals, even at the workplace, as can be seen in Table~\ref{tab:knowanduse}. As with any emerging disruptive technology, the high level of awareness can be due to the media's extensive coverage, which might have played a significant role in shaping people's perceptions of ChatGPT's trustworthiness and future impacts. 

The present study was designed to determine the effect of user characteristics on the perceived trustworthiness of AI and NLG tools, especially chatbots, addressing \textbf{RQ1}. The analysis of the results show that different sets of users have different adoption rates and willingness to use and trust NLG tools, as reported in Section~\ref{sec:rq1}. Complimentarily, the perception of machine-written text characteristics such as fluency greatly depends on user background and usage of these tools. 
The second major finding was that participants, in average, could not distinguish machine-generated text from human-written text, which answers \textbf{RQ2}, discussed in Section~\ref{sec:rq2}. The main weakness of this study was the paucity of participants in the survey. Notwithstanding the relatively limited sample, this finding is consistent with that of (i) Salminen et al. who stated that human raters exhibit low accuracy in detecting machine-generated reviews for products~\cite{Salminen2022}, (ii) Köbis2021 et al. who stated that people cannot differentiate AI-generated from human-written poetry~\cite{Köbis2021}, and (iii) Schuster et al. who reported the limited accuracy of participants in evaluating short machine-generated sentences~\cite{Schuster2020}, among others.

In turn, Section~\ref{sec:rq3} offers valuable insights into the concerns of NLG users from different fields, which support the work of other studies in this area linking NLG tools with misinformation and plagiarism~\cite{Ghosal2022, Tan2023} or potential loss of jobs for human translators~\cite{Jiao2023,Tan2023}. Differences in the foreseen positive and negative impacts of NLG tools also emerged from the analysis of the open-ended questions, addressing \textbf{RQ3}. 
Overall, the analysis of the responses of the open-ended questions shows Group 2 is more aware of the limitations of NLG and specific ethical concerns, particularly regarding privacy and explainability. Meanwhile, participants from Group 1 claimed to be more cautious about using NLG tools and expressed concerns about their impact on daily life. Gender bias, together with other ethical issues, were highlighted by participants to motivate the need for regulations to ensure the ethical use of these tools. 

Together, these results highlight the importance of informing AI development and deployment and addressing these issues by design to ensure that NLG tools are created, operated, and maintained ethically to benefit society as a whole. 

\subsection{Future work}

While NLP tools may not always produce accurate and unbiased outputs, they can still be helpful in reducing the workload of professionals in different fields. For instance, one application of NLP tools for linguists could be generating initial versions of real-time speech-to-text transcription or simultaneous translations: two tasks where trustworthiness is highly relevant. The output produced by the tool could then be refined and improved under human supervision by letting an operator edit, trim, or complement the necessary details to adapt the message to the receiver. In line with the recommendations of the EU AI Act, it would be the responsibility of the human to ensure that the text achieves the appropriate levels of politeness and cultural appropriateness~\cite{AIact}.

Further research is needed to gain a deeper understanding of the role of bias in NLG tools, especially those based in foundation models, and to develop strategies and governance mechanisms to mitigate it. A natural progression of this work would therefore be to analyse (i) the ways in which gender bias manifests in NLG tools, particularly those that are based on foundation models, or (ii) focus on prompt engineering with a focus on generating gender bias free content. For that, larger randomised controlled trials, with more emphasis on the text generation (i.e., prompt engineering), could provide more definitive evidence regarding the users perception of gender bias in NLG text.

To conclude, for the debate to be moved forward, verifiable information about AI and NLG tools needs to be made accessible to diverse societal actors. Unfortunately, the AI paradigm is switching to foundation models, very difficult to access for researchers, who are key in evaluating their potential harms and disseminating information about them. Therefore, one might ask oneself whether it is possible to set the ground for governance and lasting recommendations for AI design and deployment but...  \textit{How To Learn To Stop Worrying And Love the AI}?


\bibliographystyle{ACM-Reference-Format}
\bibliography{main}



\end{document}